# Approximate Probabilistic Neural Networks with Gated Threshold Logic

O. Krestinskaya[1] and A. P. James[1]

[1]Nazarbayev University, Astana, Kazakhstan, email: apj@ieee.org

*Abstract*— Probabilistic Neural Network (PNN) is a feed-forward artificial neural network developed for solving classification problems. This paper proposes a hardware implementation of an approximated PNN (APNN) algorithm in which the conventional exponential function of the PNN is replaced with gated threshold logic. The weights of the PNN are approximated using a memristive crossbar architecture. In particular, the proposed algorithm performs normalization of the training weights, and quantization into 16 levels which significantly reduces the complexity of the circuit.

## I. INTRODUCTION

Probabilistic Neural Network (PNN) in comparison to the other neural network types have a faster training phase where the training data is used for computing the probability density functions (PDF) of each data category. The basic PNN consists of four main layers: input pattern, summation, and output layer. The output of the $i$–th pattern layer is $g_i = (1/\sqrt{2\pi\sigma^2}) \times exp((xW_i - 1)/\sigma^2)$, where $x$ is a layer input, $W_i$ is a corresponding weight and $\sigma$ is a smoothing factor [1]. In the summation layer, the outputs of the neurons of the same class are summed up, and in the output layer the weighted summation is performed to determine the input class. PNN performs the faster training and classification, however, memory requirement for storing weight values is large because each class has separate set of corresponding neurons in all the PNN layers.

Simplifying the PNN architecture when there is an increase in number of classes, features, and samples is a challenging problem. In the past, attempts have been made to modify PNN [2], however, due to the complexity of the algorithms and exponential part of the activation function that are complex to realize accurately using analog or digital signal processing circuits, the hardware implementation of PNN has not been proposed yet. Comparing to the software implementation, the hardware design ensures an increase in processing speed and reduction in power consumption of the classification system. In this paper, we propose a novel design of PNN, namely approximate PNN (APNN) with gated threshold logic. The exponential component of the algorithm is replaced by the threshold logic based decision making process, which simplifies the implementation of the activation function. The dot product multiplication operation is performed in the memristive crossbar array, which has been proven to be efficient for various neuromorhcic and neural network architectures [3,4,5,6,7,8,9,10].

## II. APPROXIMATE PROBABILISTIC NEURAL NETWORK

In APNN, the exponential probability calculation is simplified and replaced with the thresholding operation with a threshold $\theta$: $g_i = 1$, if $|\frac{xW_i}{\sigma^2} - 1| < \theta$ and $g_i = 0$, otherwise. In addition, we implement APNN algorithm with adaptive threshold, where the threshold of each class is adapted during the training stage and each class has a unique threshold values.

This modification of PNN simplifies the design of analog hardware implementation for the network. The proposed design of APNN is shown in Fig. 1. The probabilistic weights corresponding to the training stage are stored in the crossbar as a resistance of the memristor. The weights for each separate class are stored in a separate memristive crossbar consisting of GST memristors [11]. Each column of the crossbar corresponds to a particular training sample; therefore, the number of columns in a crossbar represents the number of training samples for a certain class. The number of crossbar rows refers to the number of features in the data. For example, if the network is tested for 3 classes with 10 training samples in each class and each sample consists of 4 features, the network should include 3 separate crossbars with 4 rows and 10 columns. Therefore, each memristor in separate column stores a particular value of the probability for a single training sample. As the GST memristors are used the probabilities are quantized to 16 levels. Fig. 2 shows the quantization levels used in the proposed design corresponding to the voltage levels of probabilities.

The memristor crossbar architecture is used to implement dot product operation and the current from each crossbar column is read sequentially. A given crossbar column is activated by the transistor $Mr$ by applying the control signal $Vc = 1V$, and the applied control signal for deactivated transistors is $Vc = 0V$. To eliminate the effect of the circuits to the crossbar, the current buffer based on a current mirror is used. The current buffer inverts the current and inverted current is applied to current-to-voltage converter (IVC). The positive output voltage from the IVC is fetched into the comparator which compares the input voltage to the threshold value. There are two possible scenarios: (1) the algorithm based on the constant threshold logic for all the classes, when all the comparators in the system are set to the same threshold voltage, and (2) the adaptive threshold logic algorithm, when the comparator in each class has a unique threshold voltage that is determined during the training stage.

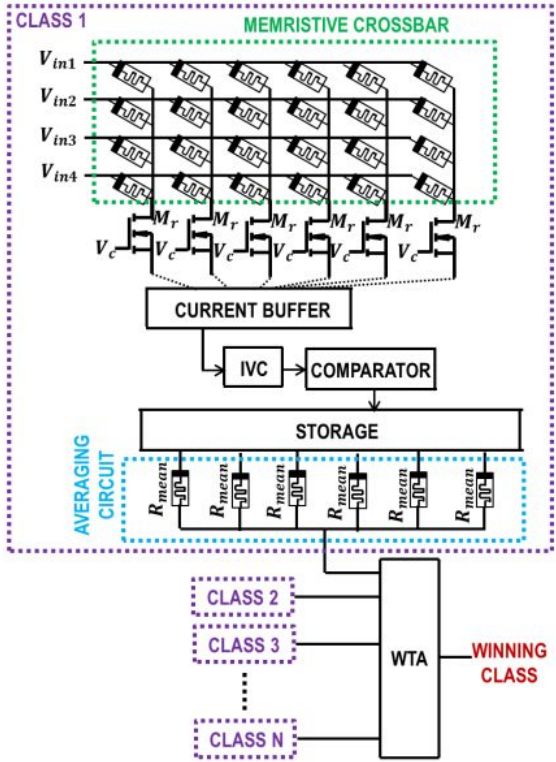

Fig.1. Overall APNN architecture.

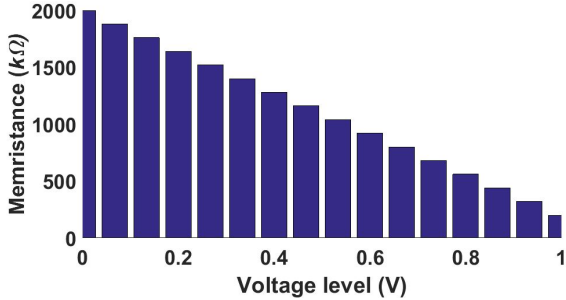

Fig.2. Quantization levels corresponding to 16-level GST memristor used in the proposed design.

The obtained values from the comparator are stored in a separate storage unit, especially if the number of training samples is large. If the number of training samples is small, the sequence of outputs can be stored using analog shift register. The benefit of such an approach is that the separate external storage unit is not required. Next, all outputs from the storage unit are fetched to the averaging circuit and the average value of the output is obtained. If the input to the crossbar is the same as the probability values stored in it, the mean of the comparator outputs is large. After this, the outputs of all crossbar corresponding to each class are compared using Winner-Takes-All (WTA) circuit shown in Fig. 3, and the class of the input pattern is determined.

Fig. 4 shows the analog circuit implementation of the separate components. Current buffer is based on standard current mirror principle and produces the negative output current to ensure that the output from IVC is positive. IVC consists of the operational amplifier with the memristors $R = 200k\Omega$. The circuit is adapted for $180nm$ CMOS technology. The transistor parameters are: $M1 = 0.18\mu/3\mu$, $M2 = 0.18\mu/30\mu$, $M3 = M4 = 0.18\mu/36\mu$, $M5 = M6 = 0.18\mu/6\mu$, $M7 = 0.18\mu/45\mu$, $M8 = M9 = M10 = 0.18\mu/18\mu$ and the capacitors $C1 = 10pF$ and $C2 = 1pF$. The threshold voltage of GST memristor used in the amplifier is greater than 1V and this allows the memristor to keep the state during the processing. The comparator circuit consists of 9 transistors: $M11 = 0.18\mu/0.72\mu$ and $M12 = 0.18\mu/0.36\mu$. $Vth$ corresponds to the threshold value of the comparator. The dependence of the required threshold value $Vth$ hand desired threshold level is shown in Fig. 5. For example, if the real threshold level for the probability is $0.4V$, it is required to set $Vth = 0.1V$.

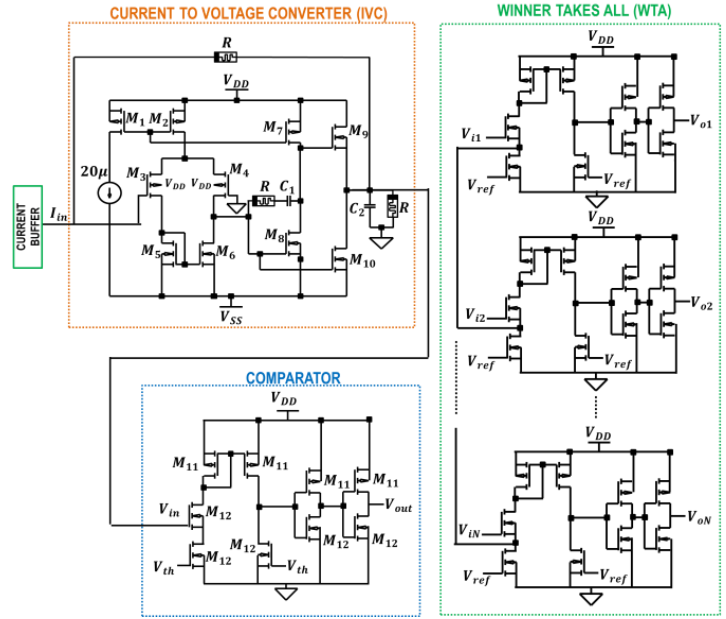

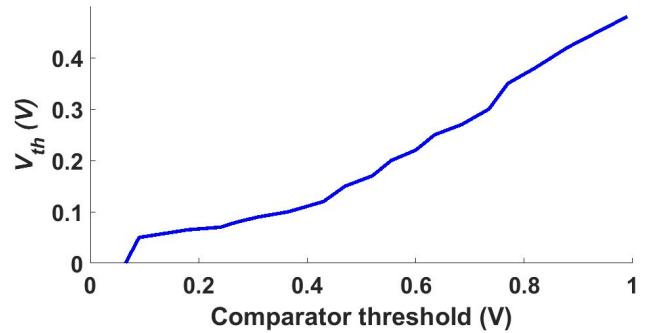

Fig.3. Circuit components used in the proposed design including current to voltage converter, comparator and Winner Takes All circuit.

Fig.4. Required comparator threshold $Vth$ mapped to the real threshold voltage.

The final stage of the processing is the class identification of the input to the system, which is performed by the WTA circuit shown in Fig. 4. WTA consists of the comparator circuits with the same parameters of the transistors as the comparator from

Fig. 4. The comparator cells are connected and the same $Vref = 0.3V$ is applied to each comparator. The input $ViN$ refers to the inputs to the WTA, where $N$ is the number of classes. The corresponding output voltages are denoted by $VoN$. The output of the comparator is $VoN = 1V$, when the classes of the input sample and the crossbar match.

## III. SIMULATION RESULTS

The proposed methods have been simulated in SPICE for TSMC 180nm CMOS technology and the full system level performance of the proposed algorithm was verified in MATLAB using IRIS database without additional preprocessing. The comparison of the conventional PNN and proposed APNN methods are shown in Table I. The proposed APNN algorithm with threshold logic outperforms the conventional PNN algorithm. When the threshold levels are quantized, the output accuracy is low because the small number of features in the database make each weight important in case of the same threshold value for all the classes. The APNN algorithm with adaptive threshold logic and quantized weights produces the maximum possible accuracy. This method is the most efficient because the threshold is adjusted for every class of the database. In addition, the algorithm can be implemented on hardware with analog circuits.

TABLE I. COMPARISON OF THE PERFORMANCE OF THE PNN METHODS.

| Method | Accuracy |
|---|---|
| Conventional PNN | 96% |
| Conventional PNN with quantized weight | 93% |
| **Approximate PNN with Threshold Logic** | **98%** |
| Approximate PNN with Threshold Logic (quantized weights) | 58.7% |
| **Approximate PNN with Adaptive Threshold Logic (quantized weights)** | **98.9%** |

The hardware simulation results for input sequence, output currents from the crossbar columns, output voltages of the corresponding IVCs, comparator output and mean calculation are shown in Fig. 5. The SPICE simulations were performed for the IRIS database with 3 classes (3 crossbars in the system), 10 training samples and 4 features in each sample corresponding to 4 input voltages $Vin1, Vin2, Vin3$ and $Vin4$. Fig. 5(a) shows the input sequence. The input sequence in time is applied in order for class 1, class 2 and class 3. The exemplar output currents from the crossbars columns are shown in Fig. 5(b). It should be notices that the current corresponding to the same class as the class of the crossbar produces the highest current, in comparison to the other inputs. Fig. 5(c) shows the output voltages of the corresponding IVCs, and Fig. 5(d) illustrates the outputs of the comparators from each crossbar. Fig. 5(e) presents the obtained mean output after reading each crossbar column. Fig. 5(f) shows the final WTA output for $Vo1, Vo2$ and $Vo3$ referring to three classes.

Table II represents the calculation of power consumption and required on-chip area of the circuit for 3 classes of inputs corresponding to 3 crossbars. The power consumption and area of the proposed circuit is *123.76 mW* and *5761.2 μm²*. The most significant amount of power and area are consumed by the operational amplifier (OpAmp) in IVC. The design can be improved by replacing the OpAmp in IVC with low-voltage OpAmp.

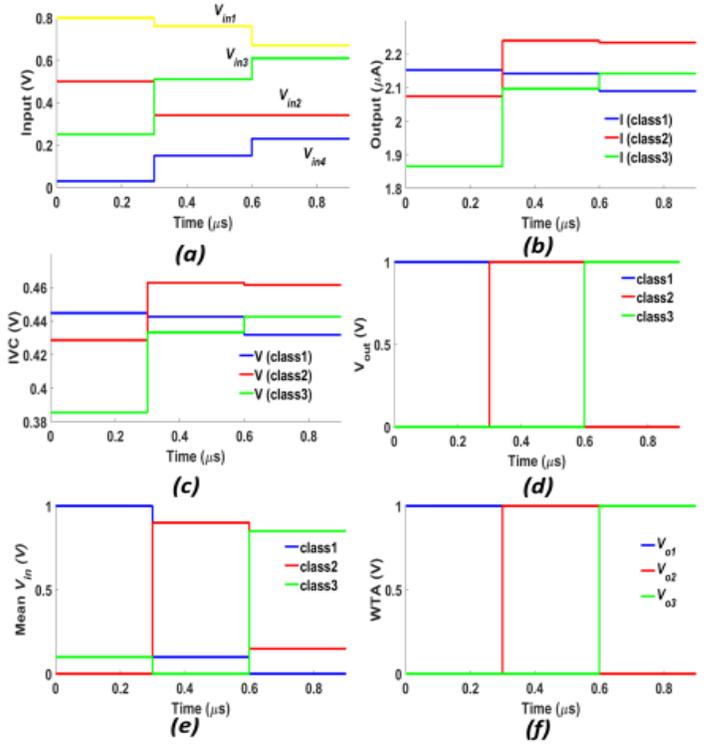

Fig.5. Timing diagram of the circuit simulation: (a) input sequence, (b) outputs of 3 different crossbars (single column example), (c) IVC out- put, (d) output of the comparators, (e) mean voltage from each crossbar and (f) WTA output.

TABLE II. POWER CONSUMPTION AND ON-CHIP AREA CALCULATION FOR THE CIRCUIT WITH 3 CLASS (3 SEPARATE CROSSBARS) AND 10 TRAINING SAMPLES PER CLASS (10 CROSSBAR COLUMNS).

| Circuit component | Power consumption | On-chip area |
|---|---|---|
| Crossbar | *5μW × 3* | *1.36μm² × 3* |
| Current buffer | *149μW × 3* | *280μm² × 3* |
| IVC | *41.1mW × 3* | *1638.7μm² × 3* |
| Comparator | *17nW × 3* | *0.5183μ² × 3* |
| WTA | *47.34pW* | *1.555μm²* |
| **Total:** | ***123.76 mW*** | ***5761.2μm²*** |

## IV. DISCUSSION

Most of the existing hardware implementations of PNN are performed on FPGA [12,13]. Comparing to this FPGA solutions, proposed analog APNN has an advantage in terms of number of components, area and power dissipation. The proposed APNN architecture can increase the processing speed, as additional elements for conversion of analog signals to digital are not required. The use of memristive elements in the design ensures the efficient weight storage, scalability and small on-chip area of the architecture.

However, there are several drawbacks and open problems that should be addressed for efficient hardware implemnentation of PNN. The analog circuit part can be further optimized to decrease power consumption and further improve the scalability of the architecture. However, the implementation of full architecture requires the design of the efficient mixed-signal control circuit to switch between the crossbar columns. Also, the efficient storage element that stores all processed outputs from the crossbar before averaging should be designed.

This paper proves that the proposed APNN architecture is useful large small scale problems, where the number of inputs is small. The application of the proposed APNN architecture for large scale problems and the inputs with large number of features is an open problem that can be addressed. In addition, the accuracy and power dissipation should be studied for large scale implementation of the architecture. Finally, the memristor related problems should be addressed. As the memristive technology is not mature [14,15,16], the effect of the real memristive devices on the overall system performance and accuracy should be studied. In addition, the selection of the appropriate memristor material and corresponding non-ideal memristor model for the simulations is also important [17,18].

## V. CONCLUSION

In this work, we implemented the analog circuits for the proposed Approximate PNN in which the conventional exponential function of the PNN was replaced with gated threshold logic. The design is based on the dot product multiplication using memristive crossbar and the weights of the probabilities of the PNN are approximated to the GST memristor levels. In particular, the proposed algorithm performs normalization of the training weights, and quantization into 16 levels which significantly reduces the complexity of the circuit. The system level simulation results and SPICE simulations demonstrated that the proposed algorithm outperforms the conventional PNN method. The result of five-fold cross-validation on Iris dataset demonstrated the accuracy of 98% for APNN algorithm with threshold logic and 98.9% for the APNN with adaptive threshold logic. The open problems include the variation of output due to instability of memristive technology and investigation of the crossbar scalability. As an extension to the reported results, we plan to include the identification of the impact of process variability, instability of the memristor switching to the APNN performance and performance evaluation of large scale system.